# Technological Devices and Their Negative Effects on Health

Vallejo-López, Alida[a], Noboa-Terán, Cesar[b], Kou-Guzmán, Juana[c]. Ramírez-Amaya, Josefina[d]

[a] **Universidad Tecnológica ECOTEC.** Facultad de Ciencias de la salud y Desarrollo Humano. Samborondón, Ecuador.

E-mail: avallejo@ecotec.edu.ec    https://orcid.org/0000-0001-7859-5268

[b] Universidad de Guayaquil. Facultad de Ciencias Médicas. Guayaquil, Ecuador.

E-mail: cesar.noboat@ug.edu.ec    https://orcid.org/0000-0002-4295-705X

[c] Universidad de Guayaquil. Facultad de Ciencias Médicas. Guayaquil, Ecuador.

E-mail: juana.koug@ug.edu.ec    https://orcid.org/0000-0002-4917-1148

[d] Universidad de Guayaquil, Guayaquil, Guayas, Ecuador.

E-mail: josefina.ramireza@ug.edu.ec    https://orcid.org/0000-0002-4338-8274

# A B S T R A C T

Technology has become a global tool that allows us to obtain information and analyze data, streamlines communication, and allows us to share images, data, videos, texts, etc. Daily activities have gone from traditional to digital. Today, it is impossible to live without an electronic device. In this context, changes in people's health observed, with various complaints ranging from visual, neurological, and concentration problems to muscular, hearing, and sleep disorders. Society must be aware of the importance of using various technological devices responsibly to protect people's health in general.

# Technological Devices and Their Negative Effects on Health

## 1. Introduction

In the new century, the world has experienced profound changes affecting human life. Without a doubt, technology has become the primary work tool. It allows us to obtain information and analyze data of all kinds. It also facilitates the process of visualizing problem solving in various branches of science and streamlines communication, allowing the sharing of images, information, data, videos, texts, etc. These significant advances allow us to optimize many processes. Daily activities have changed comparatively from previous years to the present. The inclusion of technological devices has changed the perspective of traditional tasks, transforming them into digital tasks.

Nowadays, it is impossible to live without an electronic device. Most citizens, both adults and children, frequently use electronic devices. Having a computer is essential; it is an element that gives status and places you in modernity, since it is inconceivable that someone does not have a computer, internet access, or a cell phone. These devices are characterized by being instruments that facilitate daily tasks and increase comfort in individuals' lives, such as cell phones, computers, and all those devices that serve to provide entertainment (Quintero, 2014).

This situation has altered people's habits and customs, and with it, changes in their health have been observed. (Medina & Regalado 2021), on the presence of psychosomatic problems derived from the maladaptive use of online communication technologies, which negatively affect daily activities, social relationships, physical health (poor sleep quality, obesity, impaired vision), and the psychological adjustment of adolescents.

In this regard, the presence of various discomforts is becoming evident, ranging from visual disturbances, concentration problems, muscle problems, hearing problems, and sleep disturbances. In the student environment, this situation can even compromise academic performance, causing serious distractions in young people due to various distractions, such as games, videos, social media, etc. Electronic devices affect and create problems for children if used inappropriately or for prolonged periods. In many homes, television and video games are a fundamental part of children's lives; many spend more than five hours a day in front of a screen doing nothing but interacting with these devices (UNICEF, 2017). In the workplace, these types of activities cause a loss of concentration, which affects work performance, as these distractions can lead to serious work-related accidents. This situation is generated by the impact of blue light on the nervous system. In other areas, it can affect interpersonal, emotional, and social relationships. In this context, there is a clear need for greater control over the time spent on these activities, in which children are exposed to a screen using various technological resources such as cell phones, televisions, or video game consoles. In many homes, television and video games are a fundamental part of children's lives; many spend more than five hours a day in front of a screen doing nothing but interacting with these devices (UNICEF, 2017).

## 1.1. Technological Devices

Among the technological instruments in everyday use today are cell phones, tablets, laptops, and computers.

The computer is a fundamental instrument for the development of most activities in modern society, in all areas, both personal and social, and even work. It has become a necessity, as news and all the information needed for various activities around the world are transmitted through this medium. Humans have become dependent on them for their activities, without considering the consequences of this new trend. The cell phone is an instrument that was originally created to facilitate communication between people. However, over time, other functions have been incorporated, allowing users to take photos, use files, create videos, and exchange information. It also allows them to access social networks such as Facebook, TikTok, and others. For this reason, and because of its size, it is quickly becoming the preferred instrument for many people.

**Objective:** To raise public awareness about the negative health effects that excessive use of technological devices can cause.

## 1.2. Methodology

This is a literature review study. A search conducted in the following electronic databases: Latin American and Caribbean Health Sciences Literature (LILACS), International Health Sciences Literature (MEDLINE), the Science Direct portal, and the U.S. National Library of Medicine database, 49 articles found. This is the purpose of this article. Subsequently, the remaining 23 articles read in their entirety and are included in the reflective analysis.

## 2. Alterations in Technology Users

**Neurological Disorders**

Alterations are considered present when they affect part of the nervous system; they are more serious problems and difficult to treat. Video games thought to cause photosensitive epilepsy, or PSE, a common form of epilepsy caused by repeated visual stimuli in regular patterns over a period of time and space. Another neurological problem associated with screens is migraines, caused by concentrating on a task, whether a video game or a simple Excel spreadsheet, and staring at the screen for an excessive amount of time. Its consequences are severe headaches that can last for several weeks or even become chronic.

Functional neurological disorder (NFD) refers to a neurological condition caused by changes in the way brain networks function. The National Institute of

Neurological Disorders and Stroke (NINDS) considers that anyone can develop functional neurological disorder (NFD). It is more common in women and can affect children and adults. Most people with functional movement disorders experience symptoms around the age of 20 or 30, which significantly interfere with the person's functioning and how they cope with daily life. The exact cause of FND is unknown (NIH 2025).

### 2.1. Internet Addiction Disorder

Technology obsession is a reality in the modern world and classified as a disease by specialists.

Internet Addiction Disorder (IAD): This disease caused by internet addiction, also called internet addiction (IA), is a mental health problem that involves compulsive dependence ("binge-connecting"). It is harmful to internet use, especially among young people and adolescents. Another related term is cyberaddiction, which refers to internet addiction in general, and social media addiction is a specific form of this problem. Social Media Addiction: A specific manifestation of cyberaddiction that focuses on psychological or behavioral dependence on social media platforms. According to a study on addictive behaviors on the Internet funded by the European Commission and conducted in seven European countries (Greece, Germany, the Netherlands, Iceland, Poland, Romania, and Spain), 21.3% of Spanish adolescents show signs of developing addictive behaviors on the Internet due to the time they spend online, compared to 12.7% of European averages (Robles, C 2017).

### 2.1.1. Technological insomnia.

A change that increasingly observed related to the amount of sleep that taken at night, generating a disorder called technological insomnia. This caused by the use of electronic devices for excessive periods, without any restrictions on the user. The human body has a centralized time interval known as the circadian rhythm. Sleep disturbances considered among the most serious because they do not allow for restful sleep, thus generating mental fatigue and affecting overall health. The consequences of drowsiness are often a deterioration in motor or cognitive functions, which leads to problems in interpersonal relationships and an increase in traffic and work-related accidents. (Puerto, 2015)

According to the Neurology journal Spanish Sleep Society, exposure to artificial light from technological devices causes a delay in the activity of the suprachiasmatic nucleus, altering the cycles of body functions regulated by the circadian rhythm, which, among other things, is responsible for establishing the patterns that regulate the biological variables of sleep and eating. The suprachiasmatic nucleus is a group of neurons in the medial hypothalamus that receives information from external light through the eyes. The retina contains ganglion cells with a pigment called melanopsin. This pigment transports information to the suprachiasmatic nucleus through the hypothalamic retinal tract. Information about the light-dark cycle captured, decoded, and sent to the pineal gland. This is where melatonin, a hormone that induces sleep, secreted. Therefore, its secretion is low during the day and increases at night. This process known as the circadian cycle or master clock. Artificial light has been a great advance for society, but it causes sleep disturbances, triggering health problems. The brightness of tablets, cell phones, and video game consoles disrupts the release of the necessary amount of the sleep hormone, hindering sleep, which in turn causes variations in heart rate.

### 2.1.2. Computer Vision Syndrome.

Computer vision syndrome (CVS), also known as eyestrain, is a group of visual disturbances caused by prolonged exposure (more than four hours a day) to electronic screens. It noted that vision is the process by which light transformed into nerve impulses capable of generating sensations, and that the organ responsible for performing this function is the eye. Any visual stimulus that increases the number of eye movements required for a visual task also increases neuronal effort. Nearly 60 million people worldwide experience it, resulting in lower work productivity and a lower quality of life for workers. (Vera 2022)

The level of discomfort may increase as screen time increases. The symptoms primarily associated with this syndrome: headache, blurred vision, dry eyes, neck pain, and shoulder pain (Hospital San Juan de Dios, 2016).

According to the American Optometric Association, CVS characterized by a series of visual problems derived from blue light directly entering the retina during computer use, causing visual discomfort and eye fatigue (Vallejo & Ramírez, 2023). Constantly staring at a monitor or screen for more than 4-8 hours can cause changes in people's health and performance. They may experience symptoms such as headaches, dry eyes, vision problems, constant tearing, conjunctivitis, and sleep deprivation (Pérez Tejeda, 2008).

### 2.1.3. Ergonomic Problems

The different postures people assume when using technological devices can cause physical injuries to anatomical structures in the spine, cervical, thoracic, and lumbar regions. Other anatomical structures may affected, including those corresponding to the wrist joint, which can also generate the so-called carpal tunnel syndrome, these two being the most relevant.

### 3. Results and Discussion

International studies estimate that approximately 70% of workers in the United States suffer from eyestrain due to excessive computer use, and 90% of people who use computers for three or more hours may develop this problem. Even professionals whose work depends on the processor (graphic designers, journalists, architects, accountants, and others) have been identified. Children and adolescents are also vulnerable, as they use the equipment for studies and video games. A literature review of 36 research studies on technology use in children found that delayed bedtime and reduced total sleep time are consistently associated with the use of electronic devices and social media (Cain & Gradisar, 2010). A 2008 study of adolescents in Spain revealed physical symptoms associated with ICT use, revealing that 10.1% had vision problems, 4.1% had headaches, 2.1% had joint pain, and 3.8% had back pain (Torrecillas, 2008). A similar study conducted in Santiago de Cuba during the same period revealed that 75.5% had eye fatigue and 82.2% had headaches. A 2015 study in Brazil showed that 72.1% of adolescents had neck and lower back pain, with the highest incidence being women (Fernández & García, 2010). An epidemiological profile study conducted by an Occupational Risk Administrator (ORA) showed that, in companies with more than 60 employees, 29% of them were subject to overexertion and 51% to inappropriate postures adopted by workers while performing their jobs. The incidence of some occupational diseases, including LMW, was estimated at 68,063 in 1985, and 101,645 in 2000 (Herrera and Ruiz, 2017).

Cajo et. al (2019) in their study The use of mobile electronic devices and its impact on the increase of conditions in university students, carried out on 38 students, found that "70% used Smartphones, 23% tablets and 7% laptops. 45% used it for communication, 25% used it for entertainment, 23% used it in academic activities, and 4% used it to connect to the internet and 3% used it for information. There are muscular ones with 47%, hunchback in 28%, cervical problems 11%, Whatsappitis 8% and other conditions such as Nomophobia, Phubbing, Vibranxiety and eyestrain in 7%, it concluded that the excessive use of these devices seriously influences the appearance of various conditions that affect the health of university students. "It was observed that the number of hours spent using technological devices ranged from 4 to 6 hours per day (44.4%), exposing the student community to health problems. Overall, 38.5% suffered from symptoms of eyestrain. The second most common effect was cognitive impairment, with headaches reported in 24.5% of the study.

The Garcés Orrala 2025 study provides insight into the impact of technology on the health of the student community at the Institute Technology University de Formación (ITT) who use technological devices. A sample of 100 individuals surveyed the effects of frequently using these devices for an average of 4 to 6 hours. 38.5% experienced vision problems, 24.5% headaches, while 5.2% had no symptoms, and 3.2% had sleep problems, mostly due to their daily use of electronic devices. 44.1% spent this time researching proposed activities. However, 81.5% did not receive professional help to treat the symptoms caused by the devices in question. (Garcés Orrala, J. G 2025). In the article entitled Relationship between Heart Rate Variability and the Cyclic Alternating Sleep Pattern in Patients with Insomnia, 2017, insomnia considered a pathology that damages the muscular and nervous systems, and is present in 30% of the population between 18 and 60 years of age. The study used polysomnographic recordings from five healthy subjects, five patients with psychophysiological insomnia, and five patients with insomnia without perception. The results showed that in healthy subjects, cardiac fluctuations exhibited a Brownian signal-like behavior, while an increase in the number of A1 phases observed in the electro cortical recording. The results showed a different behavior of heart rate between people with and without insomnia, consistent with the variations recorded in brain activity. The effects of this disorder often involve social, psychological, and physiological problems. Patients with insomnia observed to exhibit cardiac oscillatory patterns related to this pathology. (De León-Lomel, I, 2017).

In 2015, a study conducted at the Autonomous University of Yucatán to determine the level of sleepiness, sleep habits, and social media usage patterns among university students. Ninety-three students participated, completing a questionnaire on social media use, the Epworth Sleepiness Scale, and a sleep habits questionnaire. The results indicate differences in sleepiness levels, social media usage habits, and sleep habits across age groups, depending on the semester and gender. Similarly, a positive trade-off found between social media use and bedtime on school days. Based on total daytime computer use, the places students most frequently use to connect to the internet were home (89.2%) and school (55.9%). Students reported that their most frequently used social media sites were Facebook (96.8%), followed by YouTube (79.6%), and Twitter (71%). The participants' social media usage patterns were, on average, 4.07 hours for academic or work-related entertainment activities. (Puerto, 2015)

### 3.1. Statistics from Ecuador

In the new century, it is almost impossible to survive without an electronic device, as absolutely everything operates or connected to a wave transmission system. Therefore, whether making a payment, deposit, transaction, research, or debt, the use of computers is implicit. Humans have become very dependent on these technological devices. Even writing this article requires a computer, and publishing and accessing it will surely require one as well. This activity undoubtedly absorbs our time and part of our lives, affecting our health without us even realizing it.

In Ecuador, computer use has increased considerably both at home and at work. From 2012 to 2016, the number of laptops in households increased to 13.7 points and desktop computers to 0.3. In 2014, the country's ranking on the Technology Readiness Index (NRI), according to Telecommunications, improved, ranking 82nd out of 144 countries studied. In households, 9.8% used laptops and 24.7% used a desktop computer, according to the Ministry of Telecommunications. Smartphone ownership grew 15.2 points from 2015 to 2016; rising from 37.7% to 52.9% of people aged five and older who own an activated cell phone, according to the National Institute of Statistics and Census (INEC).

According to Internet World Stats, an organization responsible for providing internet statistics worldwide, in 2017, 13,471,736 Ecuadorians in Ecuador had internet access, meaning that three-quarters of the population used this service. A survey of 108 people revealed that 67.6% were unaware of the visual impairments caused by sitting in front of a computer screen for long periods;

32.4% said they were aware of the issue. Regarding time, 36.1% used the computer for 1 to 3 hours, 30.6% exposed for 30 minutes to 1 hour, 20.4% exposed for 3 to 6 hours, and 20.4% for more than 6 hours. 94.4% of respondents state they are aware of the damage caused by computer screens, compared to 5.6% who say they are not aware and do not consider it important. 64.8% have experienced visual discomfort. 82.4% of respondents do not know the percentage of radiation emitted by computers, compared to 17.6% who say they do. It was also learned that after reflecting on the issue, 36.1% lowered the screen brightness, 30.6% used glasses, 14.8% reduced their screen time, 12% took no preventive measures, and 3.5% used eye protection. (World Internet Statistics 2017).

**Indicators of information and communication technology**

Indicadores de la tecnología de la comunicación y la información

Fig. 1

| Grupos de edad | 2019 | 2020 | Variación significativa 2019 y 2020 |
|---|---|---|---|
| 5 a 15 años | 12,2% | 20,8% | SI |
| 16 a 24 años | 68,9% | 74,9% | SI |
| 25 a 34 años | 75,9% | 81,6% | SI |
| 35 a 44 años | 67,9% | 71,7% | SI |
| 45 a 54 años | 56,3% | 59,7% | SI |
| 55 a 64 años | 42,0% | 46,3% | SI |
| 65 y más años | 18,9% | 20,5% | No |

Fuente: Encuesta Multipropósito (2019 y 2020).

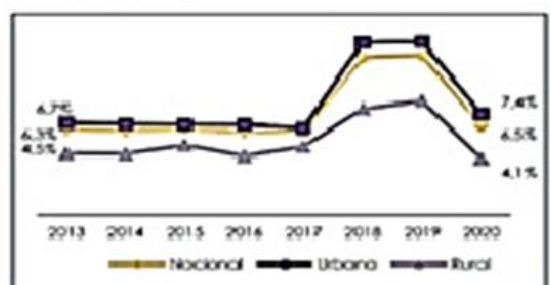

Fig. 2 — población que utiliza el internet en el trabajo, por área (2013-2020)

Notas: (1) Hasta el año 2017 se ocupa fuente: ENEMDU. Fuente: ENEMDU Diciembre (2013-2017). Encuesta Multipropósito (2018-2020).

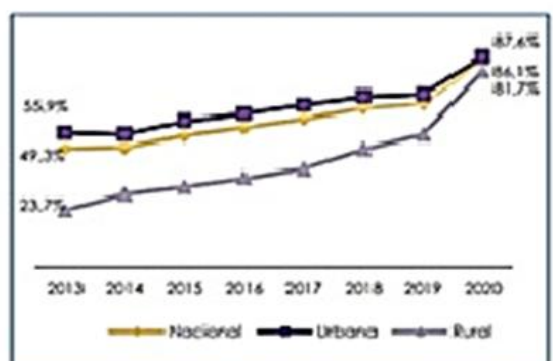

Fig. 3

Notas: (1) Hasta el año 2017 se ocupa fuente: ENEMDU. Fuente: ENEMDU Diciembre (2013-2017). Encuesta Multipropósito (2018-2020).

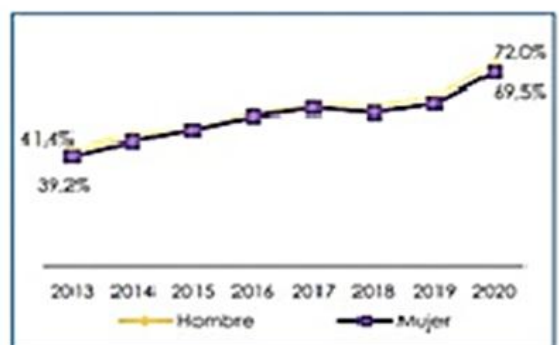

Fig. 4

Notas: (1) Hasta el año 2017 se ocupa fuente: ENEMDU. Fuente: ENEMDU Diciembre (2013-2017). Encuesta Multipropósito (2018-2020).

**Source:** INEC (National Institute of Statistics and Census 2021 Ecuadorian).

The study "Information and Communication Technology Indicators" prepared by INEC (National Institute of Statistics and Census 2021) shows an increase in the use of technological devices by age (Fig. 1) and internet access at work in

Ecuador between 2017 and 2020 (Fig. 2), as well as an increase in their use among men and women (Fig. 4).

New technologies take up so much of people's time that they stop fulfilling their obligations, restricting spaces for physical activity and replacing them with the use of electronic devices.

Excessive time spent on cell phones and computers exposes the nervous system to the impact of the energy they consume for long periods, causing dizziness, headaches, and loss of balance.

The use of technological devices exposes vision to their light, which causes uncontrolled eye damage, reduces rest time by interrupting sleep, and affects students' academic performance. The use of devices leads to incorrect postures that cause musculoskeletal problems, increases sedentary lifestyles, and causes disorders that affect people's health and promote obesity.

## 4. Conclusion

This study highlights the impact that technological devices have had on society and their negative effects on health. Therefore, it is essential to raise awareness and educate people about the amount of time they spend using technological devices in their daily activities.

**Meditate**

It recommended raise awareness among young people and users in general about the need to use technological devices only for the minimum amount of time necessary to perform activities, to avoid health complications.

**Credit authorship contribution statement**

Vallejo-López, A; Noboa-Terán, C: Writing – original draft, Supervision, Visualization, Data curation review & editing. Visualization, Validation.

Kou-Guzmán, J; Writing – original draft.

Ramírez-Amaya, J: Formal analysis, Data curation, Conceptualization. Writing – review & editing, Visualization, Validation.

**Declaration of competing interest**

The authors declare that they have no known competing financial interests or personal relationships that could have appeared to influence the work reported in this paper.

# Bibliographies

1. Cajo, B. G. H., Cajo, D. P. H., & Cajo, I. M. H. (2019). El uso de dispositivos electrónicos móviles y su impacto en el incremento de afecciones en los estudiantes universitarios. SATHIRI, 14(2), 257-269.
2. Cain, N., & Gradisar, M. (2010). Electronic media use and sleep in school-aged children and adolescents: A review. Obtenido de PubMed: https://www.ncbi.nlm.nih.gov/pubmed/20673649
   Chiza, D., Vásquez, D. & Ramírez, C. (2021). Adicción a redes sociales y ciberbullying en los adolescentes. Revista Muro de la Investigación, 6(1), 10-21. doi: 10.17162/rmi.v6i1.1437
3. DATAREPORTAL. DIGITAL 2025: EL MUNDO ESTÁ CADA VEZ MÁS CONECTADO. https://datareportal.com/reports/digital-2025-sub-section-ever-more-connected
4. DIGITAL 2025: EL MUNDO ESTÁ CADA VEZ MÁS CONECTADO https://datareportal.com/reports/digital-2025-sub-section-ever-more-connected
5. Fernández González, M. E., García Alcolea, E. E., & Martín Torres, N. (2010). Síndrome de visión de la computadora en estudiantes preuniversitarios. Revista Cubana de Oftalmología, 23, 749-757.INEC (2012). Uso del Tiempo en Ecuador. Estadístico. Quito.
6. Garcés Orrala, J. G. (2025). Impacto del uso de dispositivos electrónicos en la salud de estudiantes universitarios. CONECTIVIDAD, 6(2), 293–302. https://doi.org/10.37431/conectividad.v6i2.277
   https://revista.ister.edu.ec/ojs/index.php/ISTER/article/view/277/300
7. Herrera, C.J y Ruiz, N.A (2017). Diseño de manilla ergonómica para uso del Computador [Trabajo de Grado, Fundación Universitaria Católica – Lumen Gentium]. https://repository.unicatolica.edu.co/bitstream/handle/20.500.12237/458/FUCLG0016655.pdf?sequence=1&isAllowed=y
8. Hospital San Juan de Dios. (2016). Síndrome Visual Informático. Obtenido de https://www.sjd.es/sites/default/files/ckfinder/userfiles/files/Sindrome%20del%20Ordenador_Oftalmolog%C3%ADa.pdf
9. Kuss, D. & Griffiths, M. (2017). Social networking sites and addiction: ten lessons learned. International Journal of Environmental Research and Public Health, 14, 1-17. doi: 10.3390/ijerph14030311
10. López-López, E., Tobón, S. & Juárez-Hernández, L. (2019). Escala para evaluar artículos científicos en Ciencias Sociales y Humanidades (EACSH). Revista iberoamericana sobre calidad, eficacia y cambio en educación, 17(4), 111-125. doi: 10.15366/reice2019.17.4.006 [ Links ]
11. Medina, A., & Regalado, M. (2021). Phubbing: el otro rostro de la COVID-19. Medicina de Familia- Sociedad Española de Médicos de Atención Primaria, 47(6), 426. doi: 10.1016/j.semerg.2021.05.001
12. NIH. Instituto Nacional de Trastornos Neurológicos y Accidentes. 2025 Trastorno neurológico funcional. https://www.ninds.nih.gov/es/health-information/disorders/trastorno-neurologico-funcional
13. Pérez Tejeda AA, Acuña Pardo A, Rúa Martínez R. Repercusión visual del uso de las computadoras sobre la salud. Rev Cubana Salud Pública [revista en la Internet]. 2008 Dic [citado 2009 Ene 03];34(4): Disponible en: http://scielo.sld.cu/scielo.php?script=sci_arttext&pid=S0864-34662008000400012&lng=es

14. Puerto, M., Rivero, D., Sansores, L., Gamboa, L. y Sarabia, L. (2015). SOMNOLENCIA, HÁBITOS DE SUEÑO Y USO DE REDES SOCIALES EN ESTUDIANTES UNIVERSITARIOS. Enseñanza e Investigación en Psicología, 20 (2), 189-195.
15. Quintero, T. (2014). Prezi. Obtenido de https://prezi.com/zkd3pdaug_gq/aparatoselectronicos/
16. Robles Mora, Cristina Elizabeth (2017). Influencia del uso de los equipos tecnológicos en el rendimiento académico de los estudiantes de bachillerato general unificado de la Unidad Educativa Nacional Galápagos durante el año 2017. Trabajo de titulación previo a la obtención del Título de Licenciada en Ciencias de la Educación. Mención Informática. Carrera de Informática. Quito: UCE 133 p.
17. Torrecillas J. (2008). Uso Problemático de las Tecnologías, comunicación y el juego entre los adolescentes y jóvenes de la ciudad de Madrid. 10th ed. Salud IdAdM, Editor. Madrid: Dobleache Comunicación.
18. Unicef. (2017). Niños en un mundo digital. Obtenido de https://www.unicef.org/paraguay/spanish/UN0150440.pdf
19. Vallejo, A. B., & Ramírez, J. E. (2023). Trastornos En La Salud Visual Causados Por El Síndrome Del Computador En El Siglo Xxi.Revista Cubana de Salud Pública,49(1). Consultado el 1 de febrero de 2025, desdehttp://scielo.sld.cu/scielo.php?script=sci_abstract&pid=S0864-34662023000100003&lng=es&nrm=iso&tlng=es
20. Vera Andrade, F. N., Muñoz Flores, T. E., Rodríguez Barzola, C. V., & Gaibor Mestanza, P. M. (2022). Síndrome de ojo seco asociado al computador, manifestaciones clínicas y factores de riesgo. Sinergias Educativas. https://doi.org/10.37954/se.v0i0.98